\documentclass[12pt]{article}
\usepackage{epsfig}
\usepackage{cite}

\begin{document}
\baselineskip 0.7cm

\newcommand{\gsim}{ \mathop{}_{\textstyle \sim}^{\textstyle >} }
\newcommand{\lsim}{ \mathop{}_{\textstyle \sim}^{\textstyle <} }
\newcommand{\vev}[1]{ \left\langle {#1} \right\rangle }
\newcommand{\lsp}{ \left ( }
\newcommand{\rsp}{ \right ) }
\newcommand{\lmp}{ \left \{ }
\newcommand{\rmp}{ \right \} }
\newcommand{\llp}{ \left [ }
\newcommand{\rlp}{ \right ] }
\newcommand{\labs}{ \left | }
\newcommand{\rabs}{ \right | }
\newcommand{\EV} { {\rm eV} }
\newcommand{\KEV}{ {\rm keV} }
\newcommand{\MEV}{ {\rm MeV} }
\newcommand{\GEV}{ {\rm GeV} }
\newcommand{\TEV}{ {\rm TeV} }
\newcommand{\YR}{ {\rm yr} }
\newcommand{\mgut}{M_{GUT}}
\newcommand{\mint}{M_{I}}
\newcommand{\mgra}{M_{3/2}}
\newcommand{\mll}{m_{\tilde{l}L}^{2}}
\newcommand{\mdr}{m_{\tilde{d}R}^{2}}
\newcommand{\mllXX}[1]{m_{\tilde{l}L , {#1}}^{2}}
\newcommand{\mdrXX}[1]{m_{\tilde{d}R , {#1}}^{2}}
\newcommand{\mgy}{m_{G1}}
\newcommand{\mgl}{m_{G2}}
\newcommand{\mgc}{m_{G3}}
\newcommand{\nuR}{\nu_{R}}
\newcommand{\slL}{\tilde{l}_{L}}
\newcommand{\slLi}{\tilde{l}_{Li}}
\newcommand{\sdR}{\tilde{d}_{R}}
\newcommand{\sdRi}{\tilde{d}_{Ri}}
\newcommand{\e}{{\rm e}}
\newcommand{\bsub}{\begin{subequations}}
\newcommand{\esub}{\end{subequations}}
\newcommand{\wt}{\widetilde}
\newcommand{\tm}{\times}
\newcommand{\ra}{\rightarrow}
\newcommand{\del}{\partial}
\newcommand{\az}{a_{Z}^{}}
\newcommand{\bz}{b_{Z}^{}}
\newcommand{\cz}{c_{Z}^{}}
\newcommand{\aw}{a_{W}^{}}
\newcommand{\bw}{b_{W}^{}}
\newcommand{\dw}{d_{W}^{}}
\newcommand{\sw}{s_{W}}
\newcommand{\cw}{c_{W}}
\newcommand{\gz}{g_{Z}^{}}
\newcommand{\mz}{m_{Z}^{}}
\newcommand{\pH}{p_{H}^{}}
\newcommand{\pone}{p_{1}^{}}
\newcommand{\ptwo}{p_{2}^{}}
\newcommand{\pt}{\partial}
\newcommand{\btable}{\begin{table}[htbp]\begin{center}}
\newcommand{\etable}[1]{ \end{tabular}\caption{#1}\end{center}\end{table} }
\newcommand{\vt}{\vspace{3mm}}
\renewcommand{\thefootnote}{\fnsymbol{footnote}}
\setcounter{footnote}{1}

\makeatletter
%
%
%
%
%
\newtoks\@stequation

\def\subequations{\refstepcounter{equation}%
  \edef\@savedequation{\the\c@equation}%
  \@stequation=\expandafter{\theequation}
  \edef\@savedtheequation{\the\@stequation}
  \edef\oldtheequation{\theequation}%
  \setcounter{equation}{0}%
  \def\theequation{\oldtheequation\alph{equation}}}

\def\endsubequations{%
  \ifnum\c@equation < 2 \@warning{Only \the\c@equation\space subequation
    used in equation \@savedequation}\fi
  \setcounter{equation}{\@savedequation}%
  \@stequation=\expandafter{\@savedtheequation}%
  \edef\theequation{\the\@stequation}%
  \global\@ignoretrue}


\def\eqnarray{\stepcounter{equation}\let\@currentlabel\theequation
\global\@eqnswtrue\m@th
\global\@eqcnt\z@\tabskip\@centering\let\\\@eqncr
$$\halign to\displaywidth\bgroup\@eqnsel\hskip\@centering
     $\displaystyle\tabskip\z@{##}$&\global\@eqcnt\@ne
      \hfil$\;{##}\;$\hfil
     &\global\@eqcnt\tw@ $\displaystyle\tabskip\z@{##}$\hfil
   \tabskip\@centering&\llap{##}\tabskip\z@\cr}

\makeatother


\begin{titlepage}

\begin{flushright}
UT-02-28
\end{flushright}

\vskip 0.35cm
\begin{center}
{\large \bf Black Holes at the LHC can Determine \\ the Spin of Higgs bosons}
\vskip 1.2cm
Yosuke Uehara

\vskip 0.4cm

{\it Department of Physics, University of Tokyo, 
         Tokyo 113-0033, Japan}\\
\vskip 1.5cm

\abstract{We propose a new method to determine the
spin of Higgs bosons at the LHC by using the decay products of
black holes. Black holes may be produced if 
$\TEV$-scale gravity theories are correct, and
black holes decay into several particles. This decay results in the
emission of high energy particles, including Higgs bosons.
The difference of the degree of freedom between spin 0 and
spin 1 Higgs bosons leads to a difference of the number of 
reconstructed Higgs bosons. From this fact,
we can determine the spin of Higgs bosons with $5 \sigma$ significance
by using the integrated luminosity ${\cal L} \sim 2.4 \ {\rm fb}^{-1}$,
which can be accumulated in only one month operation.}

\end{center}
\end{titlepage}

\renewcommand{\thefootnote}{\arabic{footnote}}
\setcounter{footnote}{0}

%
%
%
%

The theories of $\TEV$-scale gravity have the possibility that
we can operate the experiments which directly access 
Planckian and transPlanckian region. 
The most exciting phenomenon in these theories may be the production of
black holes at the LHC \cite{Giddings,Dimopoulos-Landsberg,LHC}, Tevatron
\cite{Tevatron} and future linear colliders \cite{LC}. Once black holes
are produced, black holes decay by emitting the Hawking radiation.
Since we do not know the fundamental theory of the quantum gravity,
we can calculate only the decay of transPlanckian black holes and
in the following we neglect the decay processes in the Planckian region.

TransPlanckian black holes have masses $M_{BH}$, which is beyond the
fundamental scale ${\rm O}(\TEV)$. The temperature of such black
holes $T_{BH}$ is typically a few hundred $\GEV$
\cite{Giddings,Dimopoulos-Landsberg}. Thus the decay products of black holes
have energies beyond $100 \GEV$ and black holes have the possibility
to produce the particles which are not discovered yet, like Higgs bosons.
In fact, \cite{Landsberg} claimed that the black holes produced at the 
LHC decay into Higgs bosons, leading to the $5 \sigma$ discovery on
the first day of its operation. (If the fundamental scale is $1 \TEV$.)

In this letter, we show that the spin of 
Higgs bosons can also be determined by the decay of black holes.

It was difficult to determine the spin of the Higgs bosons at the LHC
Most channels suffer from too large backgrounds or 
too few events, and hence detailed studies of angular distributions
are not easy. But in this letter, we propose a completely different
method to determine the spin, not relying on the angular distributions.

As a beginning we briefly review the mechanism of black hole
production and decay.

The black hole decay occurs in several stages: balding phase,
spin-down phase, Schwarzshild phase and Planck phase \cite{Giddings}. 
Through the balding phase, black holes will 
settle down to a symmetrical rotating black holes by emitting gauge and
gravitational radiation. In the spin-down phase, by emitting the 
Hawking radiation
they lose their spin and enter Schwarzshild phase. Schwarzshild
black holes emit quanta and lose their energies. After this phase,
their properties are very difficult to analyse because of the
quantum gravity effects.

In this letter we concentrate on the non-spinning Schwarzshild
black hole. Since the production cross section of spinning black
holes is $\sim 2-3$ times higher than 
that of Schwarzshild black holes \cite{Park-Song},
and in the spin-down phase the spin-1 Higgs boson emittion is preferred
Sto the Schwarzschild phase 
because spin-1 Higgs bosons can carry angular momentum by themselves, 
making the spin of the Higgs boson easy to detect.
So our approximation is conservative.

The parton level non-spinning black hole production cross section is given by
\cite{Dimopoulos-Landsberg}:
\begin{eqnarray}
\sigma(M_{BH}) \sim \pi R_{S}^{2} = \frac{1}{M_{D}^{2}} \left[ \frac{M_{BH}}{M_{D}} (\frac{8 \Gamma(\frac{n+3}{2})}{n+2} ) \right]^{2/(n+1)}, \label{BHeq}
\end{eqnarray}
where $n$ is the number of extra dimensions, $R_{S}$ is the Schwarzshild
radius of $(n+4)$-dimensional black holes and $M_{D}$ is the fundamental
scale of $\TEV$-scale gravity theories, which is often given by:
\begin{eqnarray}
M_{D}^{n+2} V_{n} = M_{pl}^{2},
\end{eqnarray}
where $V_{n}$ is the $n$-dimensional volume of compactified extra dimensions.

But in order to validate the supossition
that the produced object is truly a black hole,
the entropy of the object must be large enough. \cite{Giddings,Cheung}
concluded that the necessary condition is
\begin{eqnarray}
M_{BH} \gsim 5 M_{D} \label{BHmin}
\end{eqnarray}

We have to integrate equation (\ref{BHeq}) in order to obtain the
total cross section. In that process we convolute the parton 
distribution function of protons with energies $\sqrt{s}/2=7 \TEV$.
The final result depends on $n$, $M_{D}$ and $M_{BH}^{min}$. From
equation (\ref{BHmin}), $M_{BH}^{min}=5 M_{D}$.

Once black holes are produced, they decay by emitting the Hawking radiation.
The decay processes of black holes 
are govened by the temperture of black holes $T_{BH}$.
The temperture is given by \cite{Myers-Perry}:
\begin{eqnarray}
T_{BH}= \frac{n+1}{4 \pi R_{S}}. \label{BHtemperature}
\end{eqnarray}
The energy spectrum of decay products is obtained by averaging
Planck formula, and it is roughly:
\begin{eqnarray}
\vev{E} \sim 2 T_{BH}.
\end{eqnarray}
So now we can obtain the number of particles emitted from
black holes $N$.
\begin{eqnarray}
N = \frac{M_{BH}}{2 T_{BH}} \propto (\frac{M_{BH}}{M_{D}})^{(n+2)/(n+1)}.
\end{eqnarray}
Black holes do not discriminate the Standard Model (SM) particles, and thus
the probability of a certain particle being emitted from them depends on
the degree of freedom of the particle. That of the SM is about 120,
and that of spin 0 Higgs bosons is 1. 

Here we end the brief review, and from now
we consider how to determine the spin of Higgs bosons by using
black holes produced at the LHC. 

First we must pay attention to the current bounds 
on $n$ and $M_{D}$. According to the review
\cite{Uehara}, the cases of $n=2,3$ are already rejected as the
candidates of the $\TEV$-scale gravity. Then the left possibility
is $4 \le n \le 7$, but since superstring theory or M-theory 
predict $10$- or $11$-dimensional spacetime, we choose $n=6$
and $n=7$ as candidates. In these cases, the constraints on $M_{D}$ 
are very weak \cite{AGASALIMIT}, and we can roughly set $M_{D}=1 \ \TEV$.

Now we have determined $n$, $M_{D}$ and $M_{BH}^{min}$. So we can
calculate the total cross section of black holes at the LHC.
It becomes \cite{Dimopoulos-Landsberg}:
\begin{eqnarray}
\sigma &\sim& 10^{6} \ {\rm fb}.
\end{eqnarray}
Here we assume the integrated luminosity ${\cal L} = 100 {\rm fb}^{-1}$.
So $\sim 10^{8}$ black holes are produced.

As stated above, the degree of freedom of the SM is about 120 and
that of Higgs bosons is 1. So as $N$ increases, the probability that
black holes decay into Higgs bosons drastically decreases. 
From the figure 1(d) of \cite{Dimopoulos-Landsberg} 
and our setup, we observe $N \ge 4$. (In the case $M_{BH}=5 \TEV$.)
Since as we increase the mass of produced black hole, the production
cross section decreases exponentially, 
we concentrate on the case of $N=4$, namely a black hole decay
into four particles, including one Higgs boson. We trigger the event
by the observation of one reconstructed Higgs boson 
and that of very high energy 
$(E_{T} \ \gsim \ {\rm a \ few} \  \TEV)$ three particles or jets. 
Three particles except one Higgs boson must satisfy the following conditions:
\begin{itemize}
 \item The possible particle candidates are gluon, quarks except top, 
       electron, muon and photon. Top, tau, $Z$ and
       $W^{\pm}$ decay before they are detected, and neutrinos and
       gravitons (including their Kaluza-Klein excitations) escape from
       the detection.
 \item The total electric charge of three particles $Q$ must satisfy
       $|Q| < 4/3$. 
\end{itemize}
The degree of freedom which satisfies these requirements becomes
roughly $110000$ d.o.f. Thus $P$, which denotes the probability that
the decay of black holes are triggered becomes:
\begin{eqnarray}
P \sim 5.1 \tm 10^{-4}.
\end{eqnarray}
So from the $10^{8}$ black holes, {\bf 51000} possible candidates are left.

We assume that the masses of Higgs bosons are $m_{h}=120 \ \GEV$. By using
a program HDECAY \cite{HDECAY}, the branching ratios of such Higgs
bosons are calculated to be:
\bsub
\begin{eqnarray}
{\rm BR} (h \ \ra \ b \ \bar{b}) &\sim& 0.66, \label{heq1} \\
{\rm BR} (h \ \ra \ W \ W^{*}) &\sim& 0.14, \\
{\rm BR} (h \ \ra \ g \ g) &\sim& 0.076, \\
{\rm BR} (h \ \ra \ \tau \ \bar{\tau}) &\sim& 0.074. \label{heq4}
\end{eqnarray}
\esub
The trigger conditions stated above are so stringent that
the processes (\ref{heq1}-\ref{heq4}) are almost background free.

We assume that in analyzing this process, Higgs boson is already
discovered. then we do not have to b-quaarks to be b-tagged,
and gluons can also be used for the reconstruction. $WW$ and $\tau \tau$
can be reconstructed when they decay hadronically or semileptonically.
So the reconstruction efficiency becomes 0.92.
So if we
operate the LHC with the integrated luminosity $100 \ {\rm fb}^{-1}$ and 
under the above trigger assumption, we can observe
$51000 \tm 0.92 =$ {\bf 47000} Higgs bosons.

Now, let us assume that Higgs bosons have spin 1. Such situations
can be realized in strongly electroweak symmetry breaking models.
If the spin of Higgs bosons is 1, the degree of freedom
becomes three, leading to the enhancement factor 3 of produced 
Higgs bosons in the decay of black holes. So in that case
we expect {\bf 140000} spin 1 Higgs bosons will be produced.

Strongly electroweak symmetry breaking models have extra
vector bosons which break electroweak symmetry and become
the alternative of the Higgs bosons in the SM. Since we are
considering $120 \GEV$ Higgs bosons, the decay of spin 1 Higgs bosons
into $WW$ is highly suppressed, and they mainly decay into
fermions. We use electrons, muons and quarks except top for
the reconstruction. Then we obtain {\bf 92000} reconstructed 
spin 1 Higgs bosons.

Now let us finalize our discussion. if the spin of Higgs bosons is 
truly 0, we can confirm the hypothesis with:
\begin{eqnarray}
\frac{|47000-92000|}{\sqrt{42000}} \sim 210 \sigma.
\end{eqnarray}
For the $5 \sigma$ discovery, only ${\cal L} = 2.4 \ {\rm fb}^{-1}$ is
needed. Low luminosity run of the LHC is designed to be 
$10^{33} \ {\rm cm}^{-2} \ {\rm s}^{-1}$, namely
$10^{-6} \ {\rm fb}^{-1} s^{-1}$. Thus only the one month operation
is enough to announce the $5 \sigma$ discovery 
of the spin 0 nature of Higgs bosons.

So black holes produced at the LHC or other colliders have the potential
not only to discover new physics \cite{Giddings,Landsberg,LC,Park-Song,Cheung},
but also to measure the properties of the SM particles more precisely 
\cite{LC,Uehara_future}. One example of the precision measurement
is just the determination of the spin of Higgs bosons.

To summarize, in this letter we propose a new
method to determine the spin of Higgs bosons at the LHC
by the decay of black holes. The method relies on the fact that
the probability a black hole decays into a particular
particle depends on the degree of freedom of the particle.
Spin 1 Higgs bosons have the freedom which is three times larger
than that of spin 0 Higgs bosons. This fact results in
the difference of the number of reconstructed Higgs bosons.
We can reject the hypothesis that Higgs bosons have spin 1 with $5 \sigma$
significance by using the integrated luminosity 
${\cal L}=2.4 \ {\rm fb}^{-1}$,
which can be accumulated in one month operation.

{\bf Acknowledgment}

Y.U. thank Japan Society for the Promotion of Science for financial
support.

\vt

\end{document}